\documentclass[prb,aps,twocolumn,showpacs,nofootinbib]{revtex4}
\usepackage{graphicx}
\usepackage{amsmath}
\usepackage{epstopdf}
\usepackage{amsbsy}
\usepackage{color}

\usepackage{bm} 

\begin{document}

\title{Coulomb Disorder Effects on ARPES and NQR Spectra in Cuprates}
\author{Wei Chen$^{1}$, Giniyat Khaliullin$^{2}$, and Oleg P. Sushkov$^{1}$}
\affiliation{$^{1}$School of Physics, University of New South Wales, 
Sydney 2052, Australia \\
$^{2}$Max-Planck-Institut f$\ddot{u}$r Festk$\ddot{o}$rperforschung, 
Heisenbergstrasse 1, D-70569 Stuttgart, Germany}

\date{\today}

\begin{abstract}
The role of Coulomb disorder, either of extrinsic origin or introduced by
dopant ions in undoped and lightly-doped cuprates, is studied. 
We demonstrate that charged surface defects in an insulator lead to
a Gaussian broadening of the Angle-Resolved Photoemisson Spectroscopy (ARPES) 
lines. The effect is due to the long-range nature of the Coulomb interaction.
A tiny surface concentration of defects about a fraction of one per cent 
is sufficient to explain the line broadening observed in  
Sr$_2$CuO$_2$Cl$_2$, La$_2$CuO$_{4}$, and Ca$_{2}$CuO$_{2}$Cl$_{2}$.
Due to the Coulomb screening, the ARPES spectra evolve dramatically with 
doping, changing their shape from a broad Gaussian form to narrow 
Lorentzian ones. To understand the screening mechanism and the 
lineshape evolution in detail, we perform Hartree-Fock simulations with 
random  positions of surface defects and dopant ions. To check validity 
of the model we calculate the Nuclear Quadrupole Resonance (NQR) lineshapes 
as a function of doping and reproduce the experimentally observed NQR 
spectra. Our study also indicates opening of a substantial
Coulomb gap at the chemical potential. 
For a surface CuO$_2$ layer the value of the gap is of the order 
of 10 meV while in the bulk it is reduced to the value about a few meV. 

\end{abstract}

\pacs{
74.72.Dn, 
79.60.Ht, 
76.60.Gv, 
73.20.At 
}
\maketitle

\section{Introduction}
During last two decades, the Angle-Resolved Photoemission Spectroscopy (ARPES) 
has developed to be one of the most important methods in the physics of 
strongly correlated systems.\cite{Damascelli03} Although the mechanism and 
physics behind the method is well understood, there are still issues remaining 
open to date. The large ARPES linewidth observed in insulating parent compounds 
Sr$_2$CuO$_2$Cl$_2$, La$_2$CuO$_{4}$, and Ca$_{2}$CuO$_{2}$Cl$_{2}$ is one of 
such problems. Already in the first experiments with Sr$_2$CuO$_2$Cl$_2$,
\cite{wells95,Poth97} it has been demonstrated that while the 
quasiparticle dispersion is well described by the extended t-J model,
\cite{Toh00,sushkov97} the quasiparticle width about 0.4 eV is difficult 
to reconcile with the predictions of this model alone. 
Similar broad spectra were later observed in 
La$_2$CuO$_{4}$~\cite{Ino2000,Yoshida03,Yoshida07} and in
Ca$_{2}$CuO$_{2}$Cl$_{2}$~\cite{Shen04,Shen07}. Interestingly, this broadening 
is not quite universal: while the linewidth in Sr$_2$CuO$_2$Cl$_2$ and 
La$_2$CuO$_{4}$ is about 0.40-0.45~eV, it is only about 0.25~eV in 
Ca$_{2}$CuO$_{2}$Cl$_{2}$, possibly indicating an extrinsic origin of
the effect. Another important observation was made in 
Ref.~\onlinecite{Shen04}: the quasiparticle lines display a Gaussian 
shape which is difficult to understand in terms of quasiparticle 
damping resulting typically in lineshapes of the Lorentzian form.

Evolution of the ARPES lineshapes with doping has also been studied
intensively.\cite{Ino2000,Yoshida03,Yoshida07,Shen04,Shen07} At doping as 
small as $3\%$, the lineshape has already changed dramatically: a narrow 
peak of a Lorentzian shape is found to emerge from a shoulder-like broad 
background, with the intensity roughly proportional to doping. 

The doping dependence of the $^{63}$Cu Nuclear Quadrupole 
Resonance (NQR) spectrum in La$_{2-x}$Sr$_x$CuO$_{4}$\cite{imai93,Singer02} 
displays a totally different behavior. The NQR line is very narrow in the 
parent compound; the effect of doping is to broaden and shift the spectra 
to higher frequency. Since NQR is a local probe of hole 
density, the broad spectrum indicates a very inhomogeneous hole density 
profile in the bulk of the sample, which is a consequence of the intrinsic 
disorder due to random La $\to$ Sr substitutions~\cite{Singer02}.

An explanation of the broad ARPES lines in undoped cuprates was suggested 
in Ref.~\onlinecite{Shen04}. According to this scenario, the strong 
interactions between holes and optical phonons lead to the
Franck-Condon broadening of the spectral functions. A detailed numerical study 
of a single hole in the t-J model coupled to optical 
phonons~\cite{Mishchenko04} has confirmed
that by appropriate tuning of the hole-phonon coupling one can properly 
reproduce the ARPES spectra (see also Ref.~\onlinecite{Gunnarsson08} for 
a review). In the Franck-Condon broadening picture, multiple phonon subbands
are generated by a photoexcited hole and the observed broad 
ARPES line corresponds to the hole-phonon incoherent background. The narrow 
quasiparticle line still exists, but it is practically invisible due to 
strongly suppressed quasiparticle residue. 

While the electron-phonon mechanism enhanced by correlation effects in Mott
systems \cite{Mishchenko04,Gunnarsson08} is able to explain  broad ARPES
lines in insulating cuprates, some questions remain to be clarified. 
As noticed in Ref.~\onlinecite{Prelovsek06}, strong suppression of 
quasiparticle peak due to Franck-Condon mechanism implies also a drastic 
enhancement of a hole effective mass from its "bare value" 
$m^* \approx 2m_e$ calculated within the t-J model. The resulting large 
mass polarons are then readily trapped by defects, 
e.g., by a negatively charged Sr dopant ion due to Coulomb attraction.  
For such a strong localization on atomic scales, the antiferromagnetic 
order would survive up to a very high doping level (similar to the case of 
Zn substitution), which contradicts the experimental data. 
In fact, the hole localization length in a lightly doped 
La$_{2}$CuO$_{4}$ ($x \leq 0.01$) is known to be about 10$\AA$ corresponding
to a moderate mass $m^* \approx 2m_e$ 
(see Refs.~\onlinecite{Chen91,Kastner98}). 
Recent calculations~\cite{filipps07} suggest that nonlocal nature of
electron-phonon interaction and longer-range hoppings may help to 
resolve the above difficulties. To reach a conclusive picture, however,
further theoretical studies of the electron-phonon mechanism in cuprates 
at {\it finite density} of holes are required. 

In the present paper we consider a different mechanism for broadening of the 
ARPES spectra. The mechanism is based on Coulomb disorder effects. 
There are two distinct kinds of Coulomb defects under consideration. 
The first kind is related to the doping mechanism, where random La $\to$ Sr 
substitutions create (negatively charged) Coulomb defects in the bulk. 
Bulk density of these defects is equal to the Sr concentration and hence 
equal to the doping level $x$. 

The second kind of defects are surface defects. We assume that cleaving 
the crystal creates some surface Coulomb defects that are unrelated to doping, 
and each surface defect has either positive or negative elementary charge. 
The presence of surface defects (e.g., missing surface ions) is physically 
plausible, and, in fact, they are observable with a scanning tunneling 
microscope (STM)~\cite{Koh07,Han04}. Denoting density of positive 
defects by $C$, we assume that negative defects have the same density 
to fulfil the charge neutrality condition. While the value of $C$ is material 
sensitive and not known a priori, we will demonstrate that a concentration
about a fraction of one per cent is already sufficient to explain observed
ARPES broadening in undoped compounds. One may argue that the emergence 
of a narrow quasiparticle peak upon doping disfavors the disorder picture, 
since disorder is also enhanced upon doping.\cite{Mishchenko04} However,  
interactions between holes induce nontrivial screening of impurities and 
dramatically reduce the effect of disorder on the ARPES spectra at finite 
doping. We find that the screening effects lead to the onset of narrow 
quasiparticle peaks already at doping as small as 1\%. Moreover, we will 
examine the effect of hole-hole interactions
on the density of states (DOS), where we recover the well-known 
results of localization theory,\cite{SE,AAP} and discuss their
implication on transport properties of lightly doped 
La$_{2-x}$Sr$_x$CuO$_{4}$.~\cite{ando02,Ando95,Boebinger96} 
Within the same model and approximations, we also address the disorder effects 
on NQR spectra and find a good description of the experimental data. 

Structure of the paper is as follows. In section II, we consider insulating
undoped compounds and calculate effect of surface Coulomb defects on ARPES
spectra. In section III, we consider doped compounds and calculate effect of
bulk Coulomb defects on NQR spectra. Section IV highlights the effect of
interactions on the bulk DOS. Evolution of ARPES spectra with doping is
calculated in section V, where both surface and bulk Coulomb defects are
taken into account simultaneously. In the case of doped compounds 
(Sections III, IV and V) one must consider screening of the long range 
Coulomb interaction by mobile holes, which is done numerically 
by performing many-body Hartree-Fock simulations.
It is noticed that we do not consider a superconducting pairing in
the present paper. Our primary goal here is to analyse role of the Coulomb 
disorder. Therefore, we concentrate on single particle properties and on the
long range Coulomb interactions only.

\section{ARPES lineshape in an insulator: broadening by surface 
Coulomb  defects}

We first consider the ARPES spectrum in the undoped insulating case, where a
single hole is injected into the cleaved surface of, e.g., La$_{2}$CuO$_{4}$. 
Coulomb potential energy of the hole at position ${\bf r}$ due to interaction
with surface defects is
\begin{equation}
\label{U}
U({\bm r})=\sum_l\frac{eq_l}{\epsilon_{s}\sqrt{|{\bm r}-{\bm r}_l|^2+a_d^2}} \ .
\end{equation}
Here $e$ is  charge of the hole (elementary charge) and  $q_l=\pm e\times q$
is charge of the defect. Eventually we will take $q=1$, but now we keep $q$ as
a parameter for general Coulomb disorder. We assume that the defect is located
at distance $a_d$ above the CuO$_2$ plane, and ${\bm r}_l$ is the 2D position
of the defect. 

Note that we use electromagnetic units, $1/4\pi\epsilon_0=1$, and the effective
surface dielectric constant is~\cite{BT} 
\begin{equation}
\label{es}
\epsilon_{s}=\frac{1}{2}(\epsilon+1) \ ,
\end{equation}
where $\epsilon$ is effective bulk dielectric constant.
According to Ref.~\onlinecite{Chen91} the bulk dielectric constant
is slightly anisotropic $\epsilon_c\sim 30$, $\epsilon_{ab}\sim 40$.
In this case the effective bulk constant in Eq.~(\ref{es}) reads as~\cite{BT}
$\epsilon=\sqrt{\epsilon_{ab}\epsilon_c}$.
As a representative value for cuprates, we will use $\epsilon =40$ for 
the bulk dielectric constant throughout the paper. 

The lattice spacing of planar Cu's is set to be unity, $a_0\approx 3.8\AA \to
1$, and hence the concentration of surface Coulomb defects $C_-=C_+=C$ is
measured in units of the number of defects per Cu site. In the limit of low
defect concentration, the potential energy constructed by Eq.~(\ref{U}) varies
slowly as a function of position ${\bm r}$, and one can define a distribution
function $P(U)$ as the probability to find a given value of $U$. 
As $C_-=C_+$, the average value of $U$ is zero, 
${\overline {U}}=\int UP(U)dU=0$. It is instructive to calculate the root mean
square deviation from zero, $\omega_{0}^2={\overline {U^2}}=\int U^2P(U)dU$.
Squaring Eq.~(\ref{U}) and averaging over random positions of defects we find
\begin{eqnarray}
\label{rms}
\omega_{0}^2=\left(\frac{e^2}{\epsilon_{s}a_0}\right)^2 q^2 2C\int_{0}^{L}
\frac{2\pi r dr}{r^2+a_d^2} \ ,
\end{eqnarray}  
where $L$ is the long distance (infrared) cutoff.
Thus,
\begin{eqnarray}
\label{rms1}
\omega_{0}&=&V \sqrt{4\pi C \ln\left(\frac{L}{a_d}\right)} \ , \\ 
\label{rms2}
V&=&\frac{qe^2}{\epsilon_{s}a_0} \approx q \times 190meV \;,
\end{eqnarray} 
where $V$ fixes the energy scale.
For purely random distribution of Coulomb defects the infrared
cutoff $L$ is equal to the radius of the incident photon beam
(assuming that the radius is smaller than the sample size).
In principle one can also imagine that due to some reason related to the 
cleaving process there is a long distance correlation between Coulomb
defects. In this case the infrared cutoff is equal to the corresponding 
correlation length. Fortunately, the dependence of $\omega_0$ on
the infrared cutoff is very weak. For instance, $\omega_0$ for $L= 1mm$ 
differs from that for $L= 1\mu m$ only by 30\%. It is worth mentioning that 
the infrared logarithmic divergence of the integral in Eq.~(\ref{rms}) is a 
consequence of the long range nature of the unscreened Coulomb interaction. 

It is also instructive to study the spatial correlator of the potential,
$\langle U({\bm r})U(0)\rangle$. Clearly, there is a structure in the 
correlator at a distance about average separation between defects, 
$r \sim 1/\sqrt{C}$. However, the most interesting behavior is at distances 
$ 1/\sqrt{C} \ll r \ll L$. A straightforward calculation similar to (\ref{rms}) 
yields that in this regime
\begin{equation}
\label{u2}
\frac{\langle U({\bm r})U(0)\rangle}{\langle U^2(0)\rangle}
\approx \frac{\ln\left(\frac{L}{r}\right)}{\ln\left(\frac{L}{a_d}\right)} \;,
\end{equation}
thus the potential varies slowly at the scale comparable with the infrared 
cutoff.

Now we calculate the entire distribution function of the potential,
\begin{equation}
\label{df}
P(\omega) = 
\langle\delta(\omega-U({\bm r}))\rangle \ ,
\end{equation}
where $\langle ...\rangle$ denotes  averaging over the observation point or, 
alternatively, averaging over distribution of defects.
We choose to put the observation point at the origin and perform averaging
over distribution of defects, hence Eq.~(\ref{U}) becomes 
$U(0)=\sum_{i}\frac{V}{R_i}-\sum_{j}\frac{V}{R_j}$,
where $i$ enumerates positive defects and $j$ enumerates negative ones, and
$R=\sqrt{{\bm r}^2+a_d^2}$ is the distance measured in units of lattice 
spacing. From Eq.~(\ref{df}) one obtains
\begin{eqnarray}
\label{p1}
P(\omega) = \frac{1}{2\pi}\int_{-\infty}^{+\infty}dt e^{i\omega t}
\langle\prod_ie^{-i\frac{V}{R_i}t} \prod_je^{+i\frac{V}{R_j}t} \rangle \ .
\end{eqnarray}
All defects are distributed independently, therefore
\begin{equation}
\label{p2}
\langle\prod_ie^{-i\frac{V}{R_i}t} \prod_je^{+i\frac{V}{R_j}t} \rangle
=\langle e^{-i\frac{V}{R}t}\rangle_{\bm r}^{N_+}
\langle e^{i\frac{V}{R}t}\rangle_{\bm r}^{N_-} \ ,
\end{equation}
where $N_+$ and $N_-$ are total numbers of positive and negative defects
and $\langle...\rangle_{\bm r}$ denotes averaging over the defect position.
Let us denote by $N$ the total number of sites in the square lattice, therefore
\begin{eqnarray}
\label{vr}
 \langle e^{\mp i\frac{V}{R}t}\rangle_{\bm r}=1-\frac{I_{\pm}}{N}\ ,
\end{eqnarray}
where
\begin{eqnarray}
\label{vr1}
I_{\pm}=\int d^2r\left(1-\exp\left\{\mp i\frac{V}{R}t\right\}\right)\ .
\end{eqnarray}
When deriving (\ref{vr}) we keep in mind that $\frac{1}{N}\int d^2r=1$.
Hence
\begin{eqnarray}
\label{vr2}
\langle e^{\mp i\frac{V}{R}t}\rangle_{\bm r}^{N_{\pm}}=
\left(1-\frac{I_{\pm}}{N}\right)^{N_{\pm}}\to
e^{-CI_{\pm}} \ .
\end{eqnarray}
Here we have taken into account that concentration of defects is
$C=N_+/N=N_-/N$. 
Hence, Eq.~(\ref{p2}) is transformed to
\begin{eqnarray}
\label{p3}
&&\langle\prod_ie^{-i\frac{V}{R_i}t} \prod_je^{+i\frac{V}{R_j}t} \rangle
=e^{-C\left(I_++I_-\right)}\nonumber\\
&&=\exp\left\{-2C\int d^2r\left(1-\cos{\frac{Vt}{R}}\right)\right\} \ .
\end{eqnarray}
To evaluate the integral with logarithmic accuracy, we expand
$\cos{\frac{Vt}{R}}$ at $\frac{Vt}{R} \ll 1$, obtaining 
\begin{eqnarray}
\label{i1}
&&2C\int d^2r\left(1-\cos{\frac{Vt}{R}}\right)\approx 2\pi C V^2t^2
 \int_0^L\frac{rdr}{r^2+a_d^2}\nonumber\\
&&=2\pi C V^2t^2\ln\frac{L}{a_d} \ .
\end{eqnarray}
Substituting Eqs.~(\ref{i1}) and (\ref{p3}) into Eq.~(\ref{p1}) and performing
integration over $t$, we find the Gaussian distribution for the potential:
\begin{equation}
\label{Ga}
P(\omega)=\frac{1}{\sqrt{2\pi}\omega_0}e^{-\omega^2/2\omega_0^2} \ ,
\end{equation}
where $\omega_0$ is given by Eq.~(\ref{rms1}).
Hence the half width of the potential distribution is
\begin{equation}
\label{g2}
\Gamma=\sqrt{8\ln{(2)}}\omega_0=4
\sqrt{2\pi\ln{(2)} \ln\left(\frac{L}{a_d}\right)}\times \sqrt{C}\times V \ .
\end{equation}
A numerical simulation, that statistically includes 100 defect configurations 
for system of size $\frac{L}{a_d} \geq 10$, shows a remarkable consistency 
between the potential generated by Eq.~(\ref{U}) and its analytical 
distribution, Eq.~(\ref{Ga}), with the width (\ref{g2}). 
Note that one should be cautious about the numerical
value of $a_{d}$. Since it is the effective short range cutoff of the Coulomb
potential, it must also include the size of Zhang-Rice singlet which is about
one lattice constant. The precise value of $a_d$ will be discussed in the
Sec. III, but now we take $a_{d}=3.8\AA\to 1$. The width of potential
distribution in Sr$_2$CuO$_2$Cl$_2$, La$_2$CuO$_{4}$, and
Ca$_{2}$CuO$_{2}$Cl$_{2}$ can then be estimated by choosing $q=1$ and
$L=1 mm$, which results in 
\begin{eqnarray}
\label{g3}
\Gamma&=&250\  meV \ \ \ \ \ \ \mbox{at} \ \ \ C\approx0.002  \nonumber\\
\Gamma&=&450 \  meV \ \ \ \ \ \ \mbox{at} \ \ \ C\approx0.006 \ .
\end{eqnarray} 

We suggest that Eq.~(\ref{Ga}) is exactly the ARPES line broadening function 
with the half-width given by Eqs.~(\ref{g2}) and (\ref{g3}). This can be
understood as follows: In the photoemission process a single hole is
injected in the top layer of the insulator, and there are two mechanisms for
the line broadening in the presence of disorder. 
The first one is the direct scattering of the hole from individual defect,
which is the short range mechanism and therefore its contribution to 
the broadening is proportional to the first power of concentration of
defects. We expect that this mechanism is negligible because of the low 
concentration of defects. The second mechanism is due to the fact that 
different holes are injected in different parts of the sample which have 
different potentials. Spectral functions are then broadened due to the 
potential distribution (\ref{Ga}). Contribution of this 
mechanism to broadening is proportional to the square root of concentration 
of defects, and, moreover, it is logarithmically enhanced, see Eq.~(\ref{g2}).

\begin{figure}[h]
\centering
\includegraphics[width=0.5\columnwidth,clip=true]{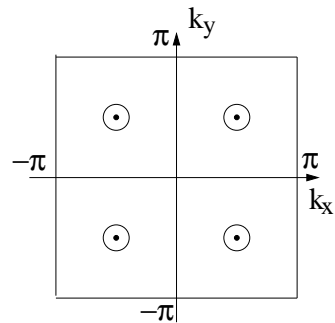}
\caption{Dispersion minima of the spinless fermion generated by Hamiltonian 
(\ref{ht}).}
\label{fig:Disp}
\end{figure}

The above picture is supported by numerical calculations that we going to
discuss now. To construct a model that can properly describe a hole motion
in undoped cuprates we first notice that there are different length scales 
in the problem: ({\it i}) The scale of the order of 1-2 lattice spacing. 
Strong correlations, such as excitations of multiple virtual magnons, 
occur at this scale. 
({\it ii}) The scale about average separation between Coulomb defects 
$\sim 1/\sqrt{C}$. Scattering from defects takes place at this scale. 
({\it iii}) The scale of $ 1/\sqrt{C} \ll r \ll L$, where logarithmically 
enhanced variations of the potential develop. Regarding the point ({\it i}), 
we do not treat here the strong correlations explicitly, 
but adopt the effective dressed hole dispersion after quantum fluctuations 
at short distances ({\it i}) are included. It is known that dispersion of the 
dressed hole has minima at points $(\pm\pi/2,\pm\pi/2)$, and is approximately
isotropic around these points~\cite{sushkov97}. The band width of the dressed 
hole is about $2J$, where $J\approx 130$ meV is the superexchange 
in the t-J model, although we do not directly employ the t-J formalism. 
Hereafter we set energy units 
\begin{equation}
\label{j}
J=130 \ meV \to 1 \ .
\end{equation}
To imitate dispersion of the dressed hole we consider spinless fermions
on a 2D square lattice. The Hamiltonian reads as
\begin{equation}
\label{ht}
H_{t}=\sum_{\langle ij \rangle}t^{\prime\prime} c^{\dag}_{i}c_{j}\;,
\end{equation}  
where $c_{i}^{\dag}$ is the hole operator at site $i$, and $t^{\prime\prime}$ 
denotes the next-next-nearest-neighbor hopping on the square lattice. 
The Hamiltonian (\ref{ht}) yields the following dispersion 
\begin{eqnarray}
\label{ht1}
\epsilon_{\bf k}=2t^{\prime\prime}(\cos 2k_{x}+\cos 2k_{y})\;.
\end{eqnarray}
The dispersion is isotropic around minima at points
$(\pm\pi/2,\pm\pi/2)$ as shown in Fig.~\ref{fig:Disp}.
We choose $t^{\prime\prime}=0.25$ to reproduce the realistic hole bandwidth as
obtained from the t-J model. Notice that in the original t-J model formalism,
there are four half-pockets inside magnetic Brillouin zone, and each pocket
has two pseudospins;\cite{sushkov97} in the present model we consider four
full pockets inside the full Brillouin zone with spinless fermions, hence the
number of ${\it charge}$ degrees of freedom is exactly the same. Clearly our
model does not have a momentum dependence of the quasiparticle residue, 
especially its suppression outside of magnetic Brillouin zone due to strong
correlations.\cite{sushkov97} However, the important point is that the Coulomb
interaction remains unchanged whatever the value of the residue is.
This is because the charge is conserved  even though holes are heavily dressed.

The hole-defect interaction due to Coulomb potential in Eq.~(\ref{U}) reads, 
after we set $J\to 1$ and $a_{0}\to 1$, as follows: 
\begin{equation}
\label{hhd}
H_{h-d}=\sum_{l,i}\frac{Q^s_l}{\sqrt{|{\bf R}_{l}-
{\bf r}_{i}|^{2}+a_d^{2}}} \; c^{\dag}_{i}c_{i}\;,
\end{equation}
with a dimensionless ''charge'' value  
\begin{equation}
\label{Q}
{Q}^s_l=\pm\frac{V}{J}\approx \pm 1.5 \ .
\end{equation} 
The superscript ``s'' stands for ``surface''.
This yields the full Hamiltonian 
\begin{equation}
H=H_{t}+H_{h-d}\;,
\end{equation}
which can be easily diagonalized on a finite size cluster where  positive 
and negative defects with concentration $C$ each are randomly distributed. 
The ARPES experiments measure the electron spectral function which can 
be calculated exactly using the cluster eigenstates and eigenenergies. 
Denoting  the hole energy as $\epsilon$ and the electron energy as $\omega$, 
we have $\omega=-\epsilon$, and hence 
\begin{eqnarray}
\label{AR}
A({\bf k},\epsilon)&=&
\sum_{n}|\langle {\bf k}|n\rangle|^{2}\delta(\epsilon-E_{n})
\nonumber \\
&=& \frac{1}{\pi}\sum_{n}\left|\sum_{i}e^{i{\bf k}\cdot {\bf r}_{i}}\alpha_{n}
({\bm r}_i)\right|^{2}\frac{\eta}{(\epsilon-E_{n})^{2}+\eta^{2}} \;, 
\nonumber \\
A({\bf k},\omega)&=&A({\bf k},-\epsilon)\;, 
\end{eqnarray}
where $\alpha_{n}({\bm r}_i)=\langle i|n\rangle$ is the coordinate
representation of the $n-$th eigenstate,
$H|n\rangle=E_{n}|n\rangle$, and $\eta=0.01$ is the artificial 
broadening of discrete energy spectrum.
Note that the artificial broadening $\eta=0.01=1.3$~meV is much smaller
than any physical contribution to the width considered in the present work. 
Therefore $\eta$ is just a technical mean to stabilize the numerical procedure
and does not contribute to physical broadening.
We perform diagonalization in a $36\times 36$ cluster with periodic boundary 
conditions, where the distance $|{\bf R}_{l}-{\bf r}_{i}|$ 
is chosen to be the shortest distance on the torus. A statistical
averaging over $100$ disorder configurations is performed.  
\begin{figure}[h]
\centering
\includegraphics[width=0.9\columnwidth,clip=true]{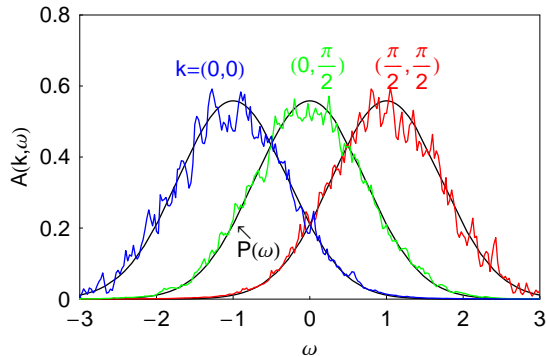}
\caption{
(Color online) ARPES spectral function (\ref{AR}) for $36\times 36$ cluster
averaged over 100 random realizations of Coulomb disorder. Concentration 
of defects is $C=0.6\%$. Values of momenta are 
${\bm k}=(\pi/2,\pi/2)$, $(0,\pi/2)$, and $(0,0)$. 
Smooth lines correspond to energy dispersion $\omega=-\epsilon_{\bf k}$ 
[with $\epsilon_{\bf k}$ given by Eq. (\ref{ht1})], broadened according to 
Eqs.~(\ref{Ga}), (\ref{g2}), and (\ref{rms2}).
In these equations, we set $L=36/\sqrt{\pi}$ and $a_d=1$.
}
\label{fig:Akw_vs_U}
\end{figure}

Fig.~\ref{fig:Akw_vs_U} shows the resulting ARPES spectral function at three
different momenta ${\bm k}=(\pi/2,\pi/2)$, $(0,\pi/2)$, $(0,0)$ together
with the analytical distribution (\ref{Ga}) of the Coulomb potential 
$P(\omega)$. In this
pedagogical example where the cluster size is certainly smaller than size
of experimental samples, we choose $C=0.6\%$. 
There is an excellent agreement
of the line shape and the line width between $P(\omega)$ and $A({\bf k},\omega)$
in all three momenta. This confirms our statement that Eq.~(\ref{Ga}) describes 
a Gaussian broadening of ARPES spectra in an insulator due to surface Coulomb 
defects, and implies that according to Eq.~(\ref{g3}), the concentration of 
$0.6\%$ surface defects (for one charge species) is sufficient to explain 
the observed broadening in La$_2$CuO$_{4}$, and $0.2\%$ is sufficient for
Ca$_{2}$CuO$_{2}$Cl$_{2}$. It is worth mentioning that this broadening is a
fairly general mechanism that works not only for Mott insulators, it is also
valid for usual band insulators.  

How one can check experimentally validity of the suggested
broadening mechanism? There are a few ``handles'' in Eq. (\ref{g2}):
the infrared cutoff $L$, the ultraviolet cutoff $a_d$, the density $C$ and
the potential $V$ of the defects. One can vary $a_d$  by probing deeper layers 
(e.g., using different photon energies): for the top
layer $a_d=1$, for the second one $a_d=4.4$, for the third layer
$a_d=7.8$, etc. Therefore the width for the second layer must be by
5-6\% smaller than the width for the top one, the width for the third 
layer is further reduced by 2-3\%. These are rather small effects. 
Another possible ``handle'' is the infrared cutoff $L$. It is natural 
to assume that $L$ is equal to the radius of the incident photon beam.
In this case focusing/defocusing of the beam changes
the width accordingly. For instance, the radius variation from $L= 1\mu m$ to
$L= 1mm$ gives rise to a sizable variation of the width by ~30\%.
It can be noticed, however, that the width is most sensitive to the amount and
the nature of defects, so systematic studies of the width variation 
as a function of the surface quality might be useful.

\section{Spectrum of Copper Nuclear Quadrupole Resonance} 

In this section we examine the effect of Coulomb disorder on the NQR
spectrum. Doping dependence of $^{63}$Cu NQR in La$_{2-x}$Sr$_x$CuO$_{4}$ has
been studied in detail in Refs.~\onlinecite{imai93} and
\onlinecite{Singer02}, which show that in the undoped parent compound the 
NQR spectrum is comprised of a very narrow line (centered at frequency 
$33.05$~MHz at T=600~K). The spectrum is shifted to 
higher frequency upon hole doping, with a line width roughly proportional to
doping. Since NQR is a local probe of real space hole distribution,
\cite{Haase04} the broad spectrum indicates a very inhomogeneous profile 
of hole density.~\cite{Singer02} 
Another interesting feature is that NQR spectra in doped
samples show a double structure: a secondary hump (the "B-line") appears 
at a frequency higher than the broad main line. The origin of the B-line
is  attributed to the Cu sites that are directly underneath the Sr 
substitutions. We will show the experimental data and compare it with our 
results later in this section. 

The NQR spectrum is obtained by calculating the hole density distribution
which is spatially nonuniform due to disorder. 
The model used in the previous section is modified here as follows. 
Since NQR is a bulk sensitive measurement the surface defects are not relevant, 
and disorder effect is solely due to the Coulomb potential of randomly 
distributed 
Sr-dopants. The total number of out-of-plane Sr ions is equal to that 
of the holes, as suggested by the doping mechanism of 
La$_{2-x}$Sr$_x$CuO$_{4}$, and each Sr-defect brings about a negative charge. 
The concentration of negative defects is therefore equal to doping, 
and no positive defects are present. Further, the strength of Coulomb 
interaction is reduced comparing to the surface case, because the bulk 
dielectric constant is larger: $\epsilon\simeq 2\epsilon_{s}$, see 
Eq.~(\ref{es}). We denote the effective dimensionless charge in bulk 
as 
\begin{eqnarray}
Q&\approx& \frac{0.5V}{J}\approx 0.75 \;,
\end{eqnarray}
which is half of the surface dimensionless charge $Q^s$, Eq.~(\ref{Q}). 
The interaction of a hole with Sr ions is then
\begin{eqnarray}
\label{hSr}
H_{h-Sr}&=&\sum_{l,i}{\cal U}_{li}
c^{\dag}_{i}c_{i}\;, \nonumber\\
{\cal U}_{li}&=&
-\frac{Q}{\sqrt{|{\bf R}_{l}-
{\bf r}_{i}|^{2}+a_d^{2}}}\;.
\end{eqnarray} 
Similarly, the Coulomb interaction between holes is described by
\begin{eqnarray}
\label{hint}
H_{int}&=&\sum_{ij}U_{ij}c^{\dag}_{i}c_{i}c^{\dag}_{j}c_{j} \ , \nonumber \\
U_{ij}&=&\frac{Q}{\sqrt{|{\bf r}_{i}-{\bf r}_{j}|^{2}+a^{2}_{d}}}\;,
\end{eqnarray}
where we use the same cutoff $a_{d}=1$ to represent the size of the 
Zhang-Rice singlet.

\begin{figure}[h]
\centering
\includegraphics[width=0.9\columnwidth,clip=true]{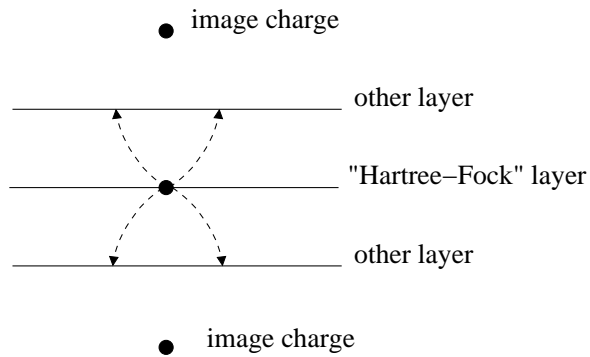}
\caption{Screening of in-plane Coulomb interaction by other layers.
Dashed lines shows electric field of  an in-plane charge bent due to a large 
polarizability of ``other layers''.
}
\label{fig:Image}
\end{figure}

We further consider the effect of multilayer screening in bulk of 
La$_{2-x}$Sr$_x$CuO$_{4}$ that contains a periodic structure of CuO$_{2}$
layers along $c$-axes. 
The electric field of a charge in a particular layer is substantially 
screened and deformed by the other layers, as shown schematically in 
Fig.~\ref{fig:Image}, due to their large polarizability.
In principle, one needs to perform a self-consistent calculation that 
includes multiple layers to account for this effect. Unfortunately, such 
a calculation is too expensive computationally. However, one can consider 
the following two limiting cases where
analytical descriptions are available. The first case is that at extremely
small doping, the polarizability of other layers is negligible, hence we
recover the single layer formulae in Eqs.~(\ref{hSr}) and (\ref{hint}). The
second case is that at sufficiently large doping, the high polarizability
implies that the electric field generated from one layer is practically
perpendicular to the surface of its nearest layers
as it is shown in Fig.~\ref{fig:Image}. In this case, one can
simply apply the method of image to account for the screening due to other
layers, by assuming that each layer is placed between two highly polarizable
media. The interactions described in Eqs.~(\ref{hSr}) and (\ref{hint}) are then
replaced by~\cite{Ho09} 
\begin{eqnarray}
\label{scr}
{\cal U}_{li}&\to&-Q\left(
\frac{1}{\sqrt{|{\bf R}_{l}-{\bf r}_{i}|^{2}+a_d
^{2}}}\right.
\nonumber\\
&&\left.+\sum_{n=1}^{\infty}\frac{2(-1)^n}
{\sqrt{|{\bf R}_{l}-{\bf r}_{i}|^{2}+(2nd)^{2}}}\right) \ ,
\nonumber\\
U_{ij}&\to&Q\left(
\frac{1}{\sqrt{|{\bf r}_{i}-{\bf r}_{j}|^{2}+a_{d}^{2}}}\right.
\nonumber\\
&&\left.+\sum_{n=1}^{\infty}\frac{2(-1)^n}
{\sqrt{|{\bf r}_{i}-{\bf r}_{j}|^{2}+(2nd)^{2}}}\right) \;,
\end{eqnarray}
where $d=13.2\AA \to 3.5$ is the separation between layers. The crossover
between these two limiting cases corresponds to the situation when the
in-plane dielectric constant due to holes is equal to the ionic dielectric
constant, which takes place at $x\approx 1\%$.\cite{Chen91} Since we are
interested in the range of $x\geq 1\%$, relevant to the NQR data discussed 
below, Eq.~(\ref{scr}) is applied to all finite doping cases in our 
simulations. Accuracy of this approximation is somewhat questionable at 
$x=1\%$, but it is fairly reasonable at higher dopings. 

\begin{figure}[h]
\centering
\includegraphics[width=0.8\columnwidth,clip=true]{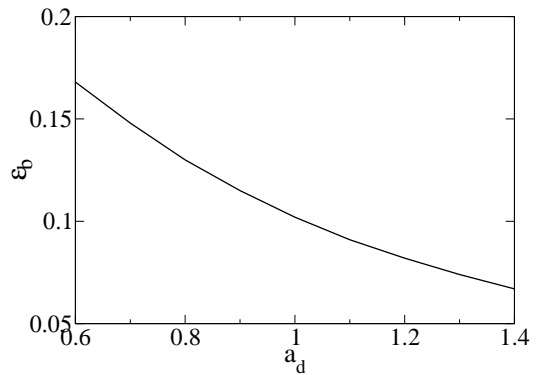}
\caption{
The hole-Sr 
binding energy as a function of the short-range cutoff 
$a_d$. We recall that we set $J=1$. The interlayer screening is taken
into account according to Eq.~(\ref{scr}).
}
\label{fig:bind}
\end{figure}

Here we give more details about the choice of $a_{d}$, under the condition
that the multilayer screening has been accounted for via Eq.~(\ref{scr}). 
The main reason for this short range cutoff is the size
of the Zhang-Rice singlet, which a priori yields $a_{d}\approx 1$. On the
other hand, the value of $a_{d}$ should properly restore the binding energy of
a single hole trapped around a Sr ion, which is known to be $\epsilon_b \sim
10 \ \mbox{meV} \sim 0.1J$.\cite{kotov07} To estimate $\epsilon_b$, we assume 
that there exists a large enough doping range where multilayer screening 
takes place via Eq.~(\ref{scr}), while the doping is still small enough 
that the in-plane hole-hole 
interaction can be ignored, and diagonalize the Hamiltonian 
$H=H_{t}+H_{h-Sr}$ with only one Sr present. The resulting binding energy 
versus $a_{d}$ is shown in Fig.~\ref{fig:bind}, where we found that
$a_{d}=1$ indeed gives the correct binding energy. For extremely low 
doping $x\ll 1\%$, we adopt the unscreened potential 
Eq.~(\ref{hSr}) without considering other layers, and found that $a_{d}=1$
gives $\epsilon_b\approx 0.23J \approx 30$ meV. This demonstrates the
importance of the multilayer screening at doping $x\geq 1\%$.
The value $a_{d}=1$ is adopted throughout this work. 

The full Hamiltonian
\begin{eqnarray}
H=H_{t}+H_{h-Sr}+H_{int}
\label{hhh}
\end{eqnarray}
can be diagonalized by the following analysis. The dimensionless 
parameter that characterizes the strength of interaction in a 2D Coulomb 
gas is\cite{Ashcroft}
\begin{equation}
\label{rs}
r_s=\frac{m^*e^2}{\epsilon \hbar^2\sqrt{\pi n}}\approx 
\frac{0.36}{\sqrt{\pi x}} \ .
\end{equation}
We see that even at $x=2\%$ the value of $r_s$ is still small $r_s\approx
1.4$. Moreover, the multilayer screening introduced in Eq. (\ref{scr}) further
reduces this value to $r_s \to 1$. Therefore, we are safely in the weak
coupling regime where the Hartree-Fock treatment is adequate. Notice that the
hole dynamics are certainly strongly correlated at the length scale about a few 
lattice spacing. These are Hubbard or t-J model correlations which result
in the dispersion of dressed holes, Eq.~(\ref{ht1}). Here the term "weak
coupling" refers to the long range Coulomb interaction between holes at 
the length scale $\geq 1/\sqrt{x}$, where the effect of the short-range
strong correlations
is already taken care of by adopting the dispersion (\ref{ht1}). The
Hartree-Fock decomposition is then applied to the hole-hole interaction 
\begin{equation}
\label{hhf}
H_{int}\to
\sum_{ij}U_{ij}\langle c^{\dag}_{i}c_{i}\rangle c^{\dag}_{j}c_{j}
-\sum_{ij}U_{ij}\langle c^{\dag}_{i}c_{j}\rangle c^{\dag}_{j}c_{i}\;.
\end{equation}
The diagonalization is again done in a $36\times 36$ cluster with periodic
boundary conditions (torus). Expectation values 
$\langle c^{\dag}_{i}c_{i}\rangle$ and $\langle c^{\dag}_{i}c_{j}\rangle$
can be calculated by
\begin{eqnarray}
\label{expe}
\langle c^{\dag}_{i}c_{i}\rangle&=&\sum_n|\alpha_n(i)|^2f(E_n) \ ,\nonumber\\
\langle c^{\dag}_{i}c_{j}\rangle&=&\sum_n\alpha_n(i)^*\alpha_n(j)f(E_n) \ ,
\end{eqnarray}
where 
\begin{equation}
\label{fd}
f(E_n)=\frac{1}{e^{(E_n-\mu)/T}-1}
\end{equation}
is the Fermi-Dirac distribution.

To determine the macroscopic chemical potential in our simulation, we adopt
the following procedure. The chemical potential at either zero or finite 
temperature in each defect configuration is determined via the charge 
neutrality condition, i.e. the number of negative defects is equal to the
number of 
holes. The average value of the chemical potential, denoted by $\mu$, is then 
calculated out of 100 defect configurations taken. We then shift the 
entire energy spectrum of each particular configuration in such a way
that the chemical  potential of the configuration is equal to this mean 
value $\mu$, which is the macroscopic chemical potential.  

\begin{figure}[h]
\centering
\includegraphics[width=0.7\columnwidth,clip=true]{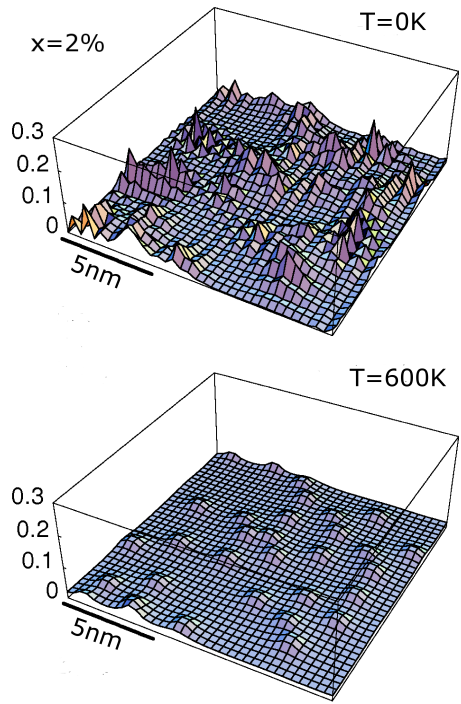}\\ 
\caption{(Color online)
Plots of the hole density in a CuO$_2$ layer deep in the bulk,
for a particular realization of random Sr positions,
at doping $x=0.02$ and two different temperatures $T=0$ and 600~K.
The length scale 5~nm ($\approx 13a_0$) is shown.
} 
\label{fig:0.02}
\end{figure}
\begin{figure}[h]
\centering
\includegraphics[width=0.7\columnwidth,clip=true]{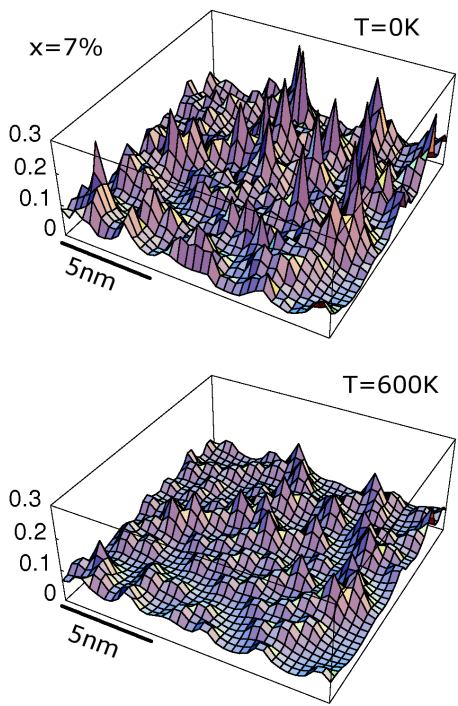}  
\caption{(Color online)
Plots of the hole density in a CuO$_2$ layer deep in the bulk,
for a particular realization of random Sr positions,
at doping $x=0.07$ and two different temperatures $T=0$ and 600~K.
}
\label{fig:0.07}
\end{figure}

Hole density plots $n_{i}=\langle c^{\dag}_{i}c_{i}\rangle$ for particular
realizations at $x=0.02$ and $x=0.07$ are shown in Figs.~\ref{fig:0.02} and
\ref{fig:0.07}, respectively, for two different temperatures 
$T=0$ and $T=600$~K. One sees very inhomogeneous density profiles, with 
a characteristic length scale of the order of a few nanometers. 
Similar nanoscale charge inhomogeneities, that closely resemble the STM 
images of underdoped cuprates 
(see, e.g., Refs.~\onlinecite{Koh07},~\onlinecite{Koh04}), have 
been also reported in previous studies~\cite{Wan01,Wan02,Zho07}. 
Interestingly, we find that increasing temperature substantially reduces 
the inhomogeneity (compare upper and lower panels in 
Figs.~\ref{fig:0.02} and \ref{fig:0.07}). This has consequences for 
NQR spectra which we address in the following.
     
\begin{figure}[h]
\centering
\includegraphics[width=1.\columnwidth,clip=true]{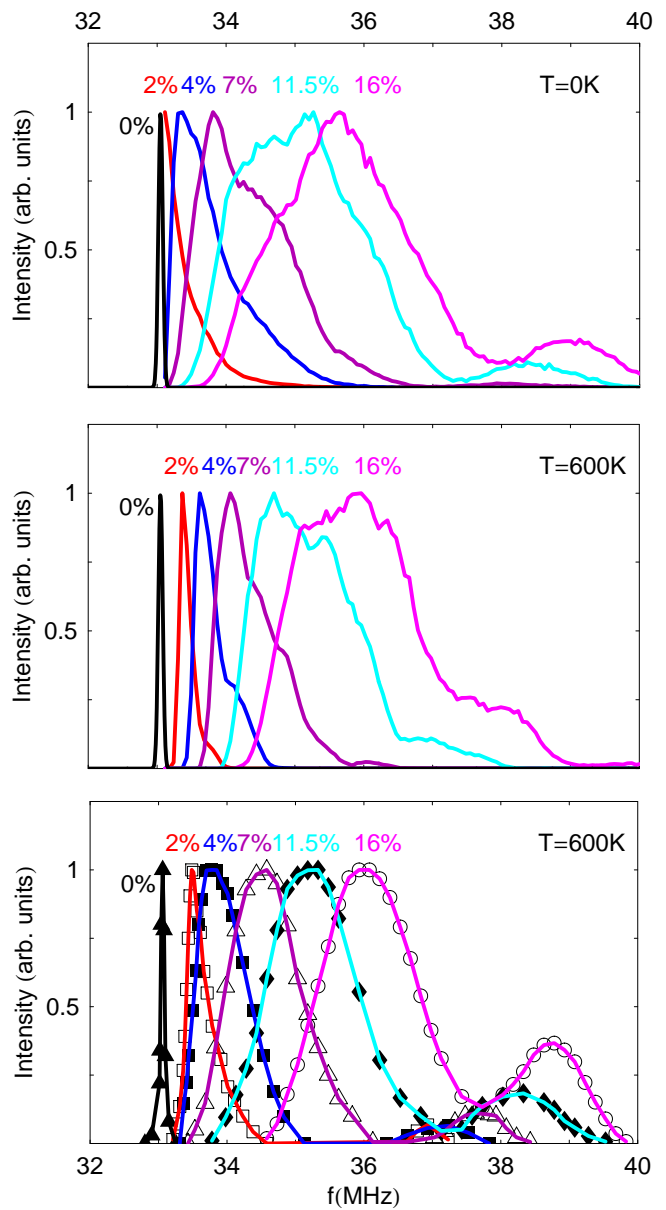}
\caption{(Color online)
The calculated (upper and middle panels) and experimental~\cite{Singer02} 
(lower panel) $^{63}$Cu NQR spectra in La$_{2-x}$Sr$_{x}$CuO$_{4}$ at
different doping levels $x$= 0.02, 0.04, 0.07, 0.115, and 0.16. The narrow 
line at $f=33.05 MHz$~\cite{imai93} corresponds to the parent compound, $x=0$. 
Apart from broadening of the spectra, doping results also in a high-frequency 
structure similar to the ``B-line'' observed in the experiment~\cite{Singer02}. 
Note that linewidth at T=600~K is smaller than that at zero temperature.   
}
\label{fig:NQRsp}
\end{figure}
The NQR frequency at a particular site $i$ is related to the hole 
density $n_{i}$ by\cite{Haase04} 
\begin{equation}
\label{fnqr}
\nu_i\approx (33.05+19\cdot n_{i}) MHz \ .
\end{equation}
Thus, the entire NQR spectrum, which effectively sums over all sites in the
sample, is proportional to the probability distribution of the hole density
${\cal P}(n)$, up to a constant shift 33.05 MHz corresponding to the
parent compound~\cite{imai93}. Calculated NQR spectra for $T=0$ and $T=600$~K
are presented in the upper and in the middle panels of Fig.~\ref{fig:NQRsp}, 
respectively. The experimental plots of Refs.~\onlinecite{Singer02} 
and \onlinecite{imai93} for La$_{2-x}$Sr$_x$CuO$_{4}$ at different dopings 
and T=600~K are  shown in the  lower panel for comparison. 
One sees that the shift and the broadening of the spectrum upon doping 
are well reproduced by the theory, with the linewidth
consistent with experimental data. 
Surprisingly, the agreement is reasonable even at 16\% doping
which is certainly too large for our low doping theory. 
The mechanism that narrows the linewidth
with temperature~\cite{Singer02} is also understood: increasing temperature 
reduces the spatial inhomogeneity of density profile, as shown in
Figs.~\ref{fig:0.02} and \ref{fig:0.07}, which results in a narrower
probability distribution ${\cal P}(n)$, and hence the narrower NQR spectrum 
(compare the upper and the middle panels in Fig.~\ref{fig:NQRsp}). 
Although our simple model does not take into account the direct action of the
Sr ion Coulomb field on the nearest Cu nuclei, which is believed to be the main 
origin of the high frequency ''B-line'',\cite{Singer02} in our numerics we 
do see a shoulder-like structure emerging at high frequency. A detailed 
investigation shows that this structure is associated with 
holes that are trapped around local potential minimum due to occasionally 
clustered $2$ to $3$ Sr defects. We suspect that these trapped holes, albeit 
not the main reason for the ''B-structure'', can have a certain contribution 
to it. The most important point here is that the main line is well understood 
in our model, which properly captures the hole density distribution 
and the screening effects in  presence of Sr defects. 

\section{Density of states and Anderson localization
in $\mbox{CuO}_2$ layer in the bulk}

In this section we address the issue of the bulk DOS in presence of disorder, 
and  relation of the DOS to the normal state dc electrical conductivity.  
It is known that La$_{2-x}$Sr$_{x}$CuO$_{4}$ exhibits the variable range hopping 
conductance at small doping $x < 0.055$;\cite{Kastner98,ando02}.
This indicates a strong localization of holes by Sr Coulomb potential.
It has been suggested that the onset of superconductivity at 
$x > 0.055$ is due to  percolation of the bound states.~\cite{sushkov05}
The hole density plots shown in Figs.~\ref{fig:0.02} and \ref{fig:0.07} 
support this suggestion: the hole
density at $x=0.02$ $(T=0)$ vanishes in large areas of the system, 
signaturing a
highly localized density profile, while the density at $x=0.07$ $(T=0)$
is nonzero
practically everywhere in the system, so wave functions are highly overlapped.

\begin{figure}[h]
\centering
\includegraphics[width=0.95\columnwidth,clip=true]{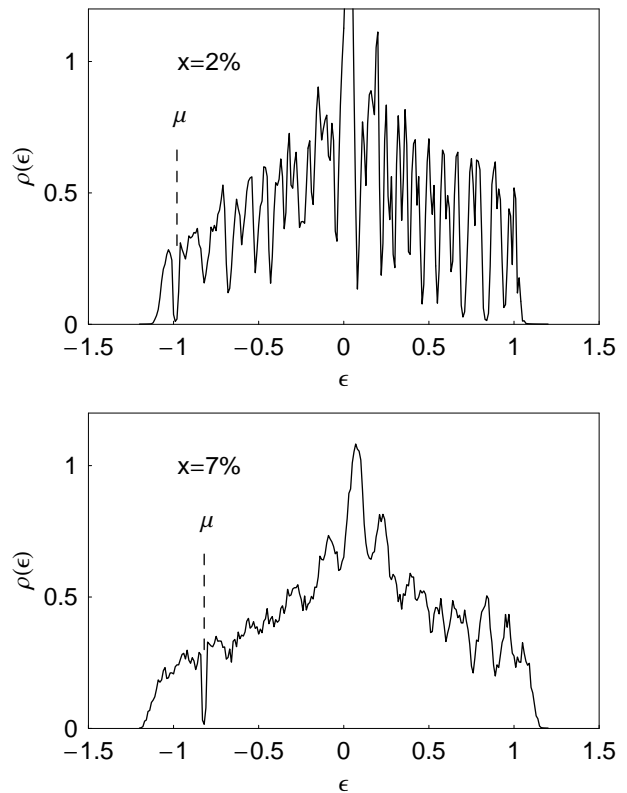}
\caption{The hole DOS in $\mbox{CuO}_2$ layer deep in the bulk
 at dopings $x=0.02$ and $x=0.07$.
DOS vanishes at the chemical potential indicating the Coulomb gap
which gradually decreases with doping.}
\label{fig:DOS}
\end{figure}

To study the problem in more detail, we have calculated the 
2D density of states, $\rho(\epsilon)$, via the standard definition:  
\begin{eqnarray}
\label{ds}
\rho(\epsilon)=\frac{1}{N}\sum_{n}\delta(\epsilon-E_{n})\;. 
\end{eqnarray}
Fig.~\ref{fig:DOS} shows the DOS calculated using the eigenenergies of 
Eq.~(\ref{hhh}) and fixing the chemical potential  
as described in the previous section.
One sees clearly a full reduction of DOS at the chemical potential
as expected by the Coulomb gap theory  in  2D systems~\cite{SE}.
The size of the gap is about $\Delta_C\sim 2.0-2.5$ meV at $x=0.02$ and it 
decreases to about $1.5-2.0$ meV at $x=0.07$.
We define $\Delta_C$  such that the total width of the gap structure in
DOS is $2\Delta_C$ as indicated in the lower panel of Fig.~\ref{DOSsurf}.
Importantly, the gap smoothly evolves through the 
percolation point $x=0.055$.  
This implies that for single particle dynamics the system remains an 
Anderson insulator even after percolation.
Probably at even larger doping, $x > 0.1$,  the Coulomb gap 
smoothly evolves to the logarithmic reduction of DOS
corresponding to the weak localization theory.\cite{AAP} 
Unfortunately, we are not able to trace this crossover because
the relatively small size of the cluster, 36$\times$36, limits our
accuracy of the gap calculation at the level $\sim 1$ meV.
The insulating behavior across the percolation point $x=0.055$
obtained in the present calculation  agrees perfectly with the 
experimental data~\cite{Ando95,Boebinger96} where the in-plane
resistivity,  measured in a very strong magnetic field that destroys 
superconductivity, shows an insulating behaviour below $\sim 50$~K for
a wide doping range up to $x \approx 15\%$.

The DOS displayed in Fig.~\ref{fig:DOS} exhibits oscillations above chemical 
potential. These oscillations is a byproduct of the finite size of the 
cluster. Maxima of the DOS correspond to degenerate
states with dispersion (\ref{ht1}) on the $36\times 36$ torus.
The oscillations must certainly disappear in the thermodynamics limit. 
However, oscillations of this kind also have an interesting physical meaning. 
In particular, the $x=0.02$ plot in Fig.~\ref{fig:DOS} indicates that while 
the quantum states 
near the chemical potential are strongly localized, the high-energy states 
well above the chemical potential are quite extended with a mean free path 
exceeding the size of the cluster used. Similarly, the smeared oscillations 
in Fig.~\ref{fig:DOS}, $x=0.07$ indicate that the mean free path 
is comparable with or less than the cluster size.

\section{Evolution of ARPES spectra with doping,
and Density of States in the Surface $\mbox{CuO}_2$ layer}

\begin{figure}[h]
\centering
\includegraphics[width=0.99\columnwidth,clip=true]{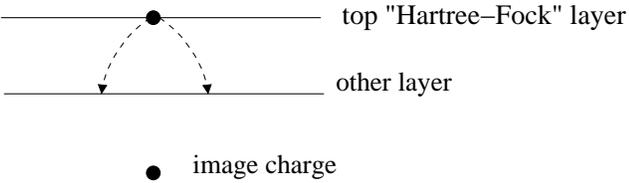}
\caption{Screening of in-plane Coulomb interaction in the top
layer by  other layers underneath.
Dashed lines shows electric field of an in-plane charge bent due to 
polarization of ``other layers''.
}
\label{fig:ImageS}
\end{figure}

To study the evolution of ARPES spectrum upon doping, we apply the above
Hartree-Fock treatment and interlayer screening picture to 
the top CuO$_2$ plane near the surface. The ARPES linewidth is determined  
now by a combined action of two types of disorder: the surface defects, 
as described by $H_{h-d}$ in Sec. II, and a randomly distributed Sr-dopant 
ions described by $H_{h-Sr}$ in Sec. III. To be specific, we fix here  
the concentration of positively/negatively charged surface defects to be 
$C_+=C_-=C=0.6\%$, as required to fit the ARPES linewidth in 
La$_{2}$CuO$_{4}$, see Eq.~(\ref{g3}). We assume that concentration 
$C$ is independent on Sr-doping since it is determined by the surface
properties unrelated to doping. Total concentration for negatively
charged defects is then $x+C=x+0.6\%$, counting both Sr-dopants and 
the negatively charged surface defects. Dimensionless charge $Q^s$ is 
again described by Eq.~(\ref{Q}), which is twice of its bulk value $Q$ 
due to the reduced dielectric constant (\ref{es}) on the surface.
Consequently, this enhances the disorder effects on the surface as we
will see below. 
The multilayer screening of the interactions on a cleaved surface is also 
different from that in the bulk, Eq.~(\ref{scr}), because even though 
other planes are still considered as highly polarizable, 
and hence the method of image is still valid for $x\geq 1\%$, the cleaved 
surface is now considered as located at a distance $d$ above a 
polarizable media, see Fig.~\ref{fig:ImageS}, 
instead of being sandwiched between two polarizable slabs. 
Collecting all these effects, we have 
\begin{eqnarray}
\label{hSrC}
H_{h-d}+H_{h-Sr}&=&\sum_{l,i}{\cal U}_{li}c^{\dag}_{i}c_{i}\;, 
\end{eqnarray}
where a disorder potential, originating either from surface or Sr defects 
located at position ${\bf R}_{l}$, is given by 
\begin{eqnarray}
{\cal U}_{li} &\to&
\pm Q^s\left(\frac{1}{\sqrt{|{\bf R}_{l}-{\bf r}_{i}|^{2}+a_d^{2}}} \right .
\nonumber\\
&&\left. \;\;\;\;
+\frac{-1}{\sqrt{|{\bf R}_{l}-{\bf r}_{i}|^{2}+(2d)^{2}}}\right) \; .
\end{eqnarray}
The difference between $H_{h-d}$ and $H_{h-Sr}$ is only in the sign of 
$Q^s$: $H_{h-Sr}$ potential is always attractive and has charge $-Q^s$, 
while its sign in $H_{h-d}$ depends on charge of the surface defect 
which can be either positive or negative. Similarly, the hole-hole 
interaction reads as 
\begin{eqnarray}
\label{hint1}
H_{int}&=&
\sum_{ij}U_{ij}c^{\dag}_{i}c_{i}c^{\dag}_{j}c_{j}\;, \\
\nonumber 
U_{ij}&\to& Q^s\left(
\frac{1}{\sqrt{|{\bf r}_{i}-{\bf r}_{j}|^{2}+a_{d}^{2}}}
+\frac{-1}
{\sqrt{|{\bf r}_{i}-{\bf r}_{j}|^{2}+(2d)^{2}}}\right). 
\nonumber
\end{eqnarray}
We then diagonalize the full Hamiltonian 
\begin{equation}
H=H_{t}+H_{h-d}+H_{h-Sr}+H_{int}
\end{equation}
following the Hartree-Fock treatment described in Eqs.~(\ref{hhf}) to 
(\ref{fd}). The procedure described in Sec. III is again applied to 
determine the macroscopic chemical potential.

\begin{figure}[h]
\centering
\includegraphics[width=0.95\columnwidth,clip=true]{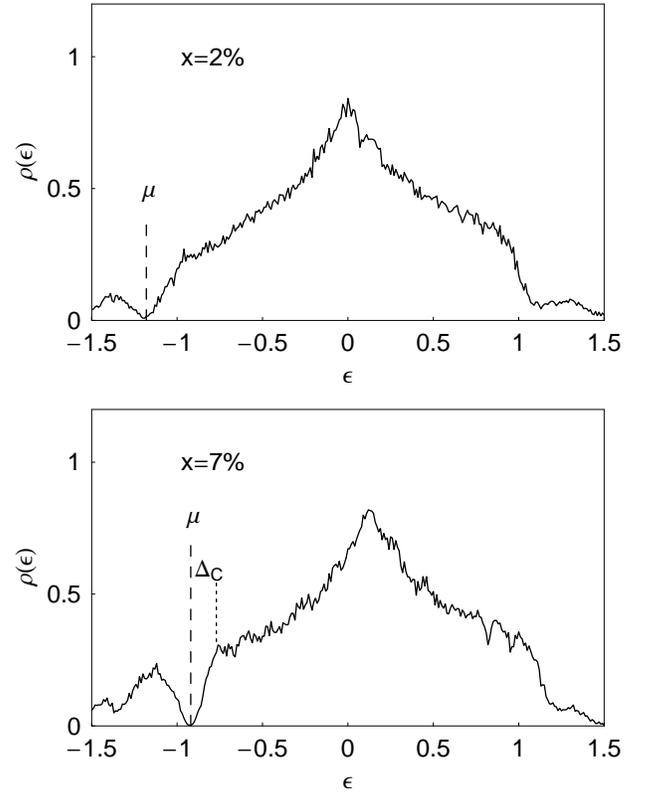}
\caption{The hole DOS in the surface $\mbox{CuO}_2$ layer
at dopings $x=0.02$ and $x=0.07$.
DOS vanishes at the chemical potential indicating the Coulomb gap
$\Delta_C$.
The size of the gap, $\Delta_C \sim 10$ meV, is larger than that in the bulk
(see Fig.~\ref{fig:DOS}).
}
\label{DOSsurf}
\end{figure}

First, we discuss DOS obtained from the numerical diagonalization. 
Fig.~\ref{DOSsurf} shows the calculated DOS for the top CuO$_2$ layer, 
where one clearly sees a Coulomb gap of the order of 
$\Delta_{C}\sim 0.1J\sim 10$~meV opening at the chemical potential.
Comparing the result with the bulk DOS shown in 
Fig.~\ref{fig:DOS}, we see a significant difference which is due to enhanced 
dimensionless charge $Q^s$: a 2D Coulomb gap scales approximately as $Q^2$,
see Ref.~\onlinecite{SE}.

There are no sizable oscillations of the DOS in 
Fig.~\ref{DOSsurf}. This means that the hole mean free path
even well above the chemical potential is much smaller than the cluster size.

Notice that values of the chemical potential on the surface, Fig.~\ref{DOSsurf},
are somewhat different from those in the bulk, Fig.~\ref{fig:DOS}.
This difference is a byproduct of our approximations. A tiny surface charging
energy and/or a tiny surface lattice deformation due to La $\to$ Sr 
substitutions can tune up the surface chemical potential from its bulk 
value. Due to these effects, which are not taken into account in the 
present model, we cannot compare the calculated chemical potential with 
experimental values. Although these 
effects can shift the chemical potential and overall energy scales, 
they do not influence the wave functions and the shape of DOS. 

\begin{figure}[htb]
\centering
\includegraphics[width=0.95\columnwidth,clip=true]{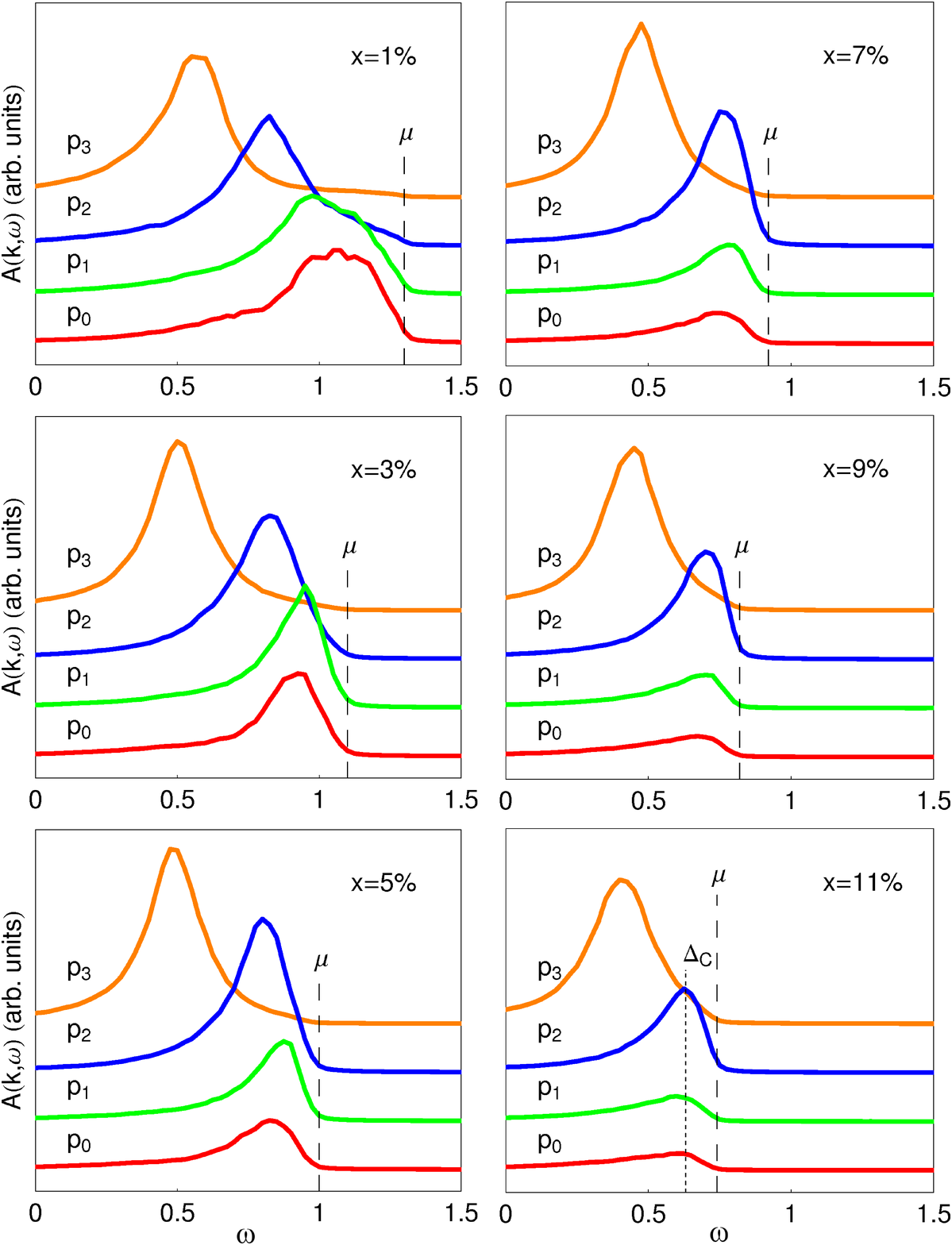}
\caption{(Color online)
The electron spectral function $A({\bf k},\omega)$ at doping levels  
$x=0.01,0.03,0.05,0.07,0.09,0.11$ calculated for momenta 
${\bf p}_0$, ${\bf p}_1$, ${\bf p}_2$, ${\bf p}_3$ specified in Eq. (\ref{pp}). 
The Coulomb gap $\Delta_{C}$ is highlighted in the $x=0.11$ plot. 
}
\label{fig:ARPESx15}
\end{figure}

Now, we turn to the ARPES spectra at finite dopings. 
The spectral functions are again calculated by using the eigenstates 
and eigenenergies given by exact diagonalization. Notice that at finite doping
and zero temperature only states above chemical potential are considered. 
This is because we are working in the hole representation, while ARPES 
spectrum is associated with the electron spectral function, therefore 
only states occupied by electrons $E_{n}>\mu$ should be summed over 
in Eq.~(\ref{AR}). It is also convenient to shift origin of the momentum 
to the dispersion minimum $(\pi/2,\pi/2)$,  
\begin{eqnarray}
\label{pk}
{\bf k}=(\pi/2,\pi/2)+{\bf p}\;,
\end{eqnarray}
and present the spectral function in terms of ${\bf p}$. 
In our model 
$A({\bf p},\omega)$ is roughly symmetric around the dispersion minimum,
$A({\bf p},\omega)\approx A(|{\bf p}|,\omega)$.
In addition, because of the $36\times 36$ finite cluster size, we can only 
calculate certain discrete values of momentum,
${\bf p}=\frac{\pi}{18}(m,n)$, where m,n are integers. On the other hand, 
the doping level $x$ can be varied continuously. In Fig.~\ref{fig:ARPESx15}, 
we present spectral functions calculated for the following momenta 
in the nodal direction
\begin{eqnarray}
\label{pp}
{\bf p}_0&=&0 \ ,\nonumber\\
{\bf p}_1&=&\left(\frac{\pi}{18},\frac{\pi}{18}\right) \ ,\nonumber\\
{\bf p}_2&=&\left(\frac{2\pi}{18},\frac{2\pi}{18}\right) \ ,\nonumber\\
{\bf p}_3&=&\left(\frac{3\pi}{18},\frac{3\pi}{18}\right) \ ,
\end{eqnarray}
and for $x=0.01\ - \ 0.11$. Surprisingly, we see very narrow lines with 
the width of the order $\Gamma \sim 0.2J \sim 30$ meV, in spite of the very 
strong disorder. This is certainly due to the Coulomb screening of both
the surface defects and Sr dopant potentials.
This residual ``small'' width $\Gamma \sim 30$~meV is due
to the hole scattering from the residual (locally unscreened) part of the
random potential. Remarkably, the residual width is quite universal,
it is practically independent of doping. On a qualitative level, this 
(somewhat unexpected) observation can be understood as a result of two  
opposite trends: On the one hand, more Sr-doping increases amount of disorder 
but, at the same time, adding more holes reduces an effective scattering
amplitude on each Sr-defect because of better screening. 
Within the present low energy effective theory, which is valid up to energies  
$E\sim J \sim 100-200$meV, the residual width is also energy/momentum 
independent.

Finally, the evolution of the small Fermi surface upon doping can be 
considered. We first notice that in the homogeneous case the
dispersion is given by Eq.~(\ref{ht1}). Hence the Fermi momentum $p_{F}$ and 
the doping $x$ are related as  
\begin{equation}
\label{pf}
x=\frac{p_F^2}{\pi} \;,
\end{equation}
if doping is small and the dispersion is roughly parabolic. Therefore, $p_F =
p_1$ at $x\approx 0.02$ and $p_F = p_2$ at $x\approx 0.08$. Interestingly, the 
relation (\ref{pf}) remains qualitatively correct even in the presence of 
strong disorder. This can be seen by the following analysis that extracts the 
Fermi momentum: We know that without disorder, the ARPES intensity 
at $p$  below $p_F$ vanishes at any finite doping, since the momentum is 
inside hole Fermi surface and no electron with this momentum can be excited. 
Therefore, the ARPES intensities at this momentum (red curves) 
in Fig.~\ref{fig:ARPESx15} are nonzero 
only because of disorder. The intensity decays very quickly when doping is 
increasing. We found that maxima of spectral functions are never exactly at 
the chemical potential, as one would expect in a system without disorder. 
Instead, we see that the maximum of each line gets closer to the chemical 
potential as doping is increased, but stops at an energy scale 
$\Delta_C\sim 0.1J\sim 10$~meV below the chemical potential. This is clearly 
due to the Coulomb gap opening in the DOS, as shown in Fig.~\ref{DOSsurf}, 
which suppresses the spectral function within 
$\mu-\Delta_{C}<\epsilon<\mu$ and shifts the maximum. 
The maximum of the $p=p_{1}$ line (green curves) approaches 
its rightmost position $\mu-\Delta_C$ at $x\approx 0.03$, so at this doping 
we say the "Fermi surface" crosses the momentum $p=p_1$. Similarly, the 
maximum of the $p=p_2$ line (blue curves) approaches its 
rightmost position at $x\approx 0.09$, hence the Fermi surface crosses 
$p=p_2$ at this doping. Following this procedure, we can identify the 
"Fermi momentum" at each doping, which can be compared with the experiments, 
although only discrete values at ${\bf p}=\frac{\pi}{18}(m,n)$ can be 
identified. 

It should be stressed that our aim here is to study the Coulomb 
disorder and its influence on the quasiparticle peak.
Strong (Hubbard or t-J model) correlations are taken into account only via
effective dispersion (\ref{ht1}) of the holon. In all other respects we 
disregard the spin degrees of freedom. Therefore, we do not 
reproduce the asymmetry of the ARPES spectral function $A({\bf p},\omega)$ 
with respect to the boundary of the magnetic Brillouin zone, and also 
strongly underestimate the ARPES intensity below 
the quasiparticle peaks, seen in the experiment as a pronounced ''hump'' 
structure. In spite of these drawbacks, the theory allows us to address 
the issue of evolution of the quasiparticle peak with disorder/doping. 
In particular, our theory explains why the ARPES lines in doped cuprates 
are relatively narrow in spite of the very strong Coulomb 
disorder.~\cite{Ino2000,Yoshida03,Yoshida07,Shen04,Shen07}

Another interesting result obtained in this section is the predicted  
Coulomb gap $\sim 10$~meV in density of states that could be observed by 
surface sensitive probes like the STM and ARPES. In fact, the gap features 
of this scale are present in the STM data for underdoped cuprates (see, e.g.,
Ref.~\onlinecite{Koh04}), and are typically attributed to the (local) 
pairing gap; the Coulomb gap might be an additional origin of these low-energy
structures. 

\section{Conclusions}

In this paper, a comprehensive study of Coulomb disorder effects in undoped 
and lightly-doped cuprates is performed, and the main results can be summarized 
as follows.  \\

\noindent
1. We have demonstrated that a very small amount of surface Coulomb 
defects leads to a dramatic broadening of ARPES spectrum in insulators. 
In particular, a concentration of defects about just a fraction of 
1\% is sufficient to explain observed ARPES line widths in
La$_2$CuO$_{4}$ and Ca$_{2}$CuO$_{2}$Cl$_{2}$. The broadened spectrum displays a 
Gaussian shape, consistent with experiments.~\cite{Shen04}
In the end of section II we have discussed possible ways to check the
suggested broadening mechanism experimentally. \\

\noindent
2. Doping process, e.g., random substitutions La $\to$ Sr  
in La$_{2-x}$Sr$_{x}$CuO$_{4}$, intrinsically creates strong inhomogeneity in 
the system. By performing Hartree-Fock calculations, we show that due to the 
strong Coulomb screening, ARPES lines obtain a very narrow width 
($\Gamma \sim 30-40$~meV) as soon as doping is higher than $\sim 1\%$, 
in spite of the very strong disorder. These results provide a natural
explanation for why the ARPES spectra undergo radical changes -- from 
very broad Gaussian to narrow quasiparticle peaks -- upon just 
a few percent doping of parent compounds.
The residual ``small'' width $\Gamma \approx 30-40$~meV is due
to the hole scattering from the residual (locally unscreened) part of the
random potential. The residual width is quite universal,
it is practically independent of doping/energy/momentum. \\

\noindent
3. The calculation of the surface density of states demonstrates that the 
top CuO$_2$ layer of La$_{2-x}$Sr$_{x}$CuO$_{4}$ is always in the
Anderson localization regime, and we predict the Coulomb gap of the order of 
$\sim 10$~meV which could be observed with STM and/or ARPES experiments. \\

\noindent
4. The calculation of the bulk density of states 
also shows  the Coulomb gap of the order of a few meV. 
The gap evolves smoothly  through the percolation point $x=0.055$.  
Hence the system remains in the Anderson localization regime,
and this explains the insulating behavior observed in transport
properties at high magnetic fields.~\cite{Ando95,Boebinger96} \\

\noindent
5. Considering Sr-doping induced disorder in La$_{2-x}$Sr$_{x}$CuO$_{4}$,  
we find a very inhomogeneous hole density profile which yields a broad 
NQR spectrum. The calculated doping and temperature dependencies 
of NQR lineshapes are consistent with experiments. \\

Altogether, the results reported here highlight a significant role played 
by Coulomb disorder effects in cuprates. In particular, screening of Coulomb 
defects (either of extrinsic origin or introduced by dopant ions) results in 
a dramatic evolution of physical properties upon doping. In this work, we 
focused mostly on the charge degrees of freedom, accounting for underlying 
magnetic correlations merely via a properly renormalized dispersion of the 
mobile holes. It remains a challenge to incorporate the magnetic degrees 
of freedom into the model explicitly, exploring thereby the coupled charge 
and spin dynamics in cuprates at short length scales.  

\section{acknowledgments}

Useful discussions with L.P. Ho, A. Fabricio Albuquerque, C.J. Hamer, 
J. Oitmaa, and T. Valla are acknowledged. We would like to thank 
P.M. Singer and T. Imai for sharing and discussion of their NQR data. 
G.Kh. thanks the School of Physics and Gordon Godfrey fund, UNSW, for kind 
hospitality. O.P.S. thanks the MPI Stuttgart for kind hospitality. 
Numerical calculation is done by the facilities of University of Florida 
High Performance Cluster.

\end{document}